\documentclass[apj]{emulateapj}

\usepackage[table,usenames,dvipsnames]{xcolor}
\usepackage{amsmath}

\begin{document}

\title{Prospects for Delensing the Cosmic Microwave Background for Studying Inflation}

\author{Gabrielle Simard, Duncan Hanson and Gil Holder}
\affil{Department of Physics, McGill University, Montreal, Quebec H3A 2T8, Canada; simardg@physics.mcgill.ca}

\begin{abstract}

A detection of excess cosmic microwave background (CMB) B-mode polarization on large scales allows the possibility of measuring not only the amplitude of these fluctuations
but also their scale dependence, which can be parametrized as the tensor tilt $n_T$.
Measurements of this scale dependence will be 
hindered by the secondary B-mode polarization anisotropy induced by gravitational lensing. 
Fortunately, these contaminating B modes can be estimated and removed
with a sufficiently good estimate of the intervening gravitational potential and
a good map of CMB E-mode polarization.  We present forecasts for how well these
gravitational lensing B modes can be removed, assuming that the lensing potential can be
estimated either internally from CMB data or using maps of the cosmic infrared 
background (CIB) as a tracer. 
We find that CIB maps are
as effective as CMB maps for delensing at the noise levels of the current generation of CMB experiments, while the CMB maps themselves will ultimately be best for delensing at polarization noise below $\Delta_P$=1 $\mu$K-arcmin.
At this sensitivity level, CMB delensing will be able to measure $n_T$ to an accuracy of 0.02 or better, which corresponds to the tensor tilt predicted by the consistency relation for single-field slow-roll models of inflation with $r=0.2$. 
However, CIB-based delensing will not be sufficient for constraining $n_T$ in simple inflationary models.

\end{abstract}


\section{Introduction}

The initial BICEP2 high-sensitivity measurements of B-mode polarization in the 
microwave sky on large angular scales ($\ell \lesssim 300$) have generated a great deal of excitement \citep{Ade:2014xna}.
This result was originally seen as a clear detection of primordial B-modes, which can be interpreted
 as the imprint of gravitational radiation from the epoch of inflation (see, for example,  \citealt{Ade:2014xna} and \citealt{Martin:2014lra}, for a review of different models).
These potential primordial B-modes could also be interpreted
 as the signature of alternatives to inflation (\citealt{Brandenberger:2014faa}; \citealt{Cai:2014xxa}; \citealt{Gerbino:2014eqa}; \citealt{Wang:2014kqa})
 or as the signature of
 topological defects (\citealt{Lizarraga:2014eaa}; \citealt{Moss:2014cra}). 
Subsequent work has shown the strong possibility that the BICEP2 signal can be explained by Galactic dust 
(\citealt{Flauger:2014qra}; \citealt{Mortonson:2014bja}; \citealt{Adam:2014bub}).
While the excess B-modes seen by BICEP2 have been confirmed by deeper measurements from the \textit{Keck Array} \citep{2015arXiv150200643A} over the same patch of sky, a joint analysis of the BICEP2, \textit{Keck Array} and \textit{Planck} maps finds that the significance of the residual signal after subtraction of the Galactic dust emission is too low to constitute a robust detection of primordial B-modes \citep{2015PhRvL.114j1301B}.

\begin{figure}[tp]
\label{fig:pspec}
\centering
\includegraphics[width=\columnwidth]{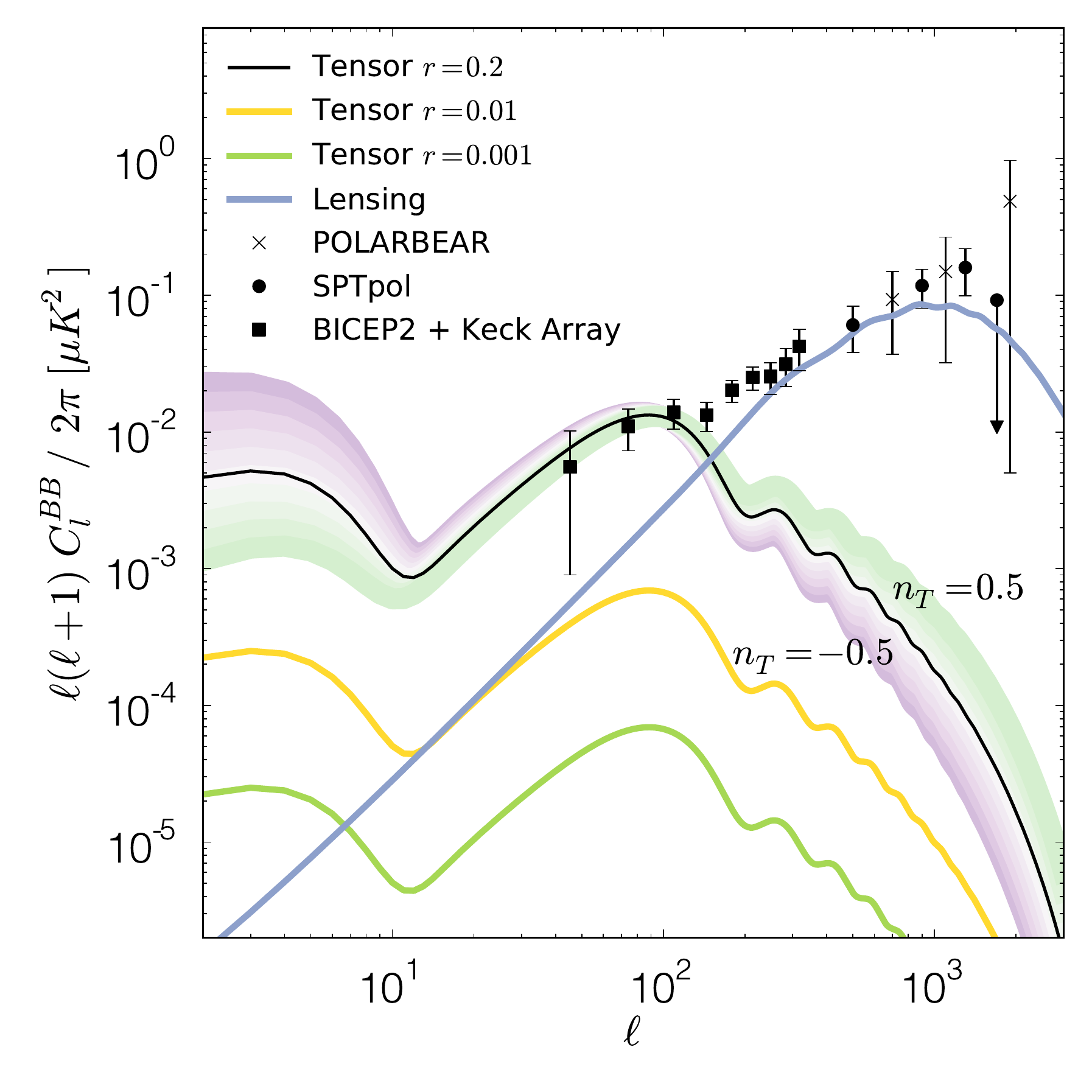} 
\caption{Theoretical power spectrum of the gravitational waves B-mode polarization and lensing B-mode polarization along with the
BICEP2/\textit{Keck Array} \citep{2015arXiv150200643A},
SPTpol \citep{2015arXiv150302315K}  and
POLARBEAR \citep{Ade:2014afa} measurements. 
The shaded region illustrates the tilt imprinted on a $r = 0.2$ primordial power spectrum from $n_T = -0.5$ (purple) to $n_T = 0.5$ (green). All tensor power spectra are for $r$ defined at $\ell=100$.}
\end{figure}

It is not yet clear whether the BICEP2 signal is due entirely to dust or has a significant primordial contribution. If the signal really is primordial, it would be interesting to measure the scale dependence of these tensor fluctuations (henceforth called ``tensor tilt'') using future polarization measurements.
Figure~\ref{fig:pspec} shows the effect of varying the tensor tilt $n_T$ on the theoretical power spectrum of the primordial B modes for a fixed value of the tensor-to-scalar ratio $r$.
The $r=0.2$ power spectra have been normalized at $\ell=100$, an angular scale well probed by the BICEP2 experiment, although for the results presented in Section \ref{sec:results} the pivot scale at which $r$ is taken will be held to $k_0 = 0.002 \, \text{Mpc}^{-1}$. 
This choice of pivot scale affects the quoted values of $r$ but not the r-marginalized constraints on $n_T$ \citep{Boyle:2014kba}.

A serious impediment to measuring the tensor tilt will be noise coming from gravitational
lensing of the cosmic microwave background (CMB) polarization fluctuations generated by scalar
fluctuations. 
Removing this signal is possible, using a procedure
known as ``delensing'' (\citealt{Kesden:2002ku}; \citealt{Knox:2002pe}).
In principle, with sufficiently high signal-to-noise measurements and in the absence of any primordial B-mode signal 
this delensing procedure can be done with arbitrarily high precision \citep{Seljak:2003pn}.
The ability to remove the lensing signal requires
a good map of the polarization fluctuations from the scalars (E-mode
polarization anisotropies) and a good estimate of the intervening 
projected gravitational potential.  E-mode polarization maps can be 
obtained only from sensitive CMB polarization measurements on the relevant
angular scales, while the gravitational potential can either be
estimated from the CMB data or obtained using astronomical sources as
tracers of the potential. 

One of the most promising CMB lensing potential tracers is the cosmic infrared background (CIB). 
The CIB is an extragalactic radiation field generated by the unresolved emission from star-forming galaxies (\citealt{Dole:2006aqw}, and references therein).
It is generated by dust which is heated by the UV light from young stars and then reradiates thermally in the  infrared with a graybody spectrum of $T\sim30K$.
Due to its higher temperature, fluctuations in the CIB dominate over the CMB on most angular scales at frequencies $\nu \gtrsim 300$GHz.
The CIB contains approximately half of the total extragalactic stellar flux, and has long been predicted to have excellent redshift overlap with the CMB lensing potential \citep{Song:2002sg}. 
Recent measurements of the cross-correlation between the lensing potential and high signal-to-noise measurements of the CIB fluctuations from the {\it Planck} and {\it Herschel} satellites have upheld these predictions (\citealt{Hanson:2013hsb}; \citealt{Holder:2013hqu}; \citealt{Ade:2013aro}; \citealt{Ade:2013hjl}).

In this paper, we evaluate prospective constraints on $n_T$ using both the CMB and the CIB to estimate the lensing potential. 
A similar study uses delensing to obtain constraints on the tensor-to-scalar ratio, for $r$ values much smaller than the BICEP2 initial claim of $r=0.2$ \citep{2015arXiv150205356S}.
We refer the reader to this work for a thorough investigation 
of the optimal use of CIB data for delensing purposes, alone or in combination with the CMB.
The paper is divided as follow: in Sect.~\ref{sec:method} we quantify the noise levels associated with these tracers of the lensing potential and E modes and describe the forecasting method used to delens the B-mode power. In Sect.~\ref{sec:results} we present the resulting constraints and our conclusions are summarized in Sect.~\ref{sec:conclusions}.

Throughout this work, we use the Planck/WP/highL fiducial cosmology \citep{Ade:2013zuv} for all cosmological parameters except $r$ and $n_T$.
We take the fiducial values of the tensor power spectrum parameters to follow the consistency relation $n_T = -r/8$,  characteristic of the single-field slow-roll inflation models \citep{1992PhLB..291..391L}.
The CMB temperature, CMB polarization and lensing potential power spectrum have been computed using the CLASS Boltzmann code \citep{blas:2011cos}.

\section{Forecasting Method}
\label{sec:method}

We discuss here the tools we used to achieve CMB delensing and to compute forecasted errors on the parameters describing the primordial B modes.
Gravitational lensing deflects CMB temperature and polarization primordial anisotropies according to \citep{Lewis:2006fu}:
\begin{align} \label{eq:remap}
T^{\text{len}}(\mathbf{\hat{n}}) \, &= T^{\text{unl}}( \mathbf{\hat{n}} +  \nabla\phi(\mathbf{\hat{n}}) )  \nonumber \\
 \left(  Q \pm iU \right)^{\text{len}}(\mathbf{\hat{n}}) \, &= \,  \left(  Q \pm iU \right)^{\text{unl}}( \mathbf{\hat{n}} +  \nabla\phi(\mathbf{\hat{n}}) ).
\end{align}
The lensing potential $\phi(\mathbf{\hat{n}})$ can be expressed as a line of sight integral:
\begin{align} \label{eq:phi_los}
\phi (\mathbf{\hat{n}}) \,=\, -2 &\int_0^{z_{\text{rec}}} \frac{dz}{H(z)} \Psi \left(z,D(z) \,  \mathbf{\hat{n}} \right)  \\
&\times \left( \frac{1}{D(z)} - \frac{1}{D(z_{\text{rec}})} \right), \nonumber
\end{align}
where $H(z)$ is the Hubble factor, $\Psi(z, \mathbf{x})$ the Newtonian gravitational potential and $D(z)$ the comoving distance to the redshift $z$. A flat universe has been assumed in writing expression (\ref{eq:phi_los}). 
Rewriting the right-hand side of Equation (\ref{eq:remap}) in the E modes and B modes CMB polarization formalism \citep{Zaldarriaga:1996xe} and expanding it up to first order in the lensing potential, we get the following approximation for the lensed B modes \citep{Hu:2000ee}:
\begin{align} \label{eq:b_lens}
B^{\, \text{len}}_{\ell m} \, \approx \, &B^{\, \text{unl}}_{\ell m} \\
& + \sum_{\ell' m'} \sum_{LM} \, f_{\ell \ell' L} \, E^{\, \text{unl}*}_{\ell' m'} 
\begin{pmatrix}
\ell & \ell' & L \\
m   &  m'  & M
\end{pmatrix}
\phi^*_{LM}. \nonumber
\end{align}
The $f_{\ell \ell' L}$ couplings are given by
\begin{equation} \label{eq:f_coupling}
f_{\ell \ell' L} \, = \, \frac{F^{-2}_{\ell \ell' L} - F^{2}_{\ell \ell' L}}{2i},
\end{equation}
where
\begin{align} \label{eq:F_coupling}
F^{s}_{\ell \ell' L} \, = & \; \frac{1}{2} \left[  -\ell(\ell + 1) + \ell'(\ell' + 1) + L(L + 1) \right]   \\
&\times \, \sqrt{ \frac{(2\ell + 1)(2\ell' + 1)(2L + 1)}{4\pi}} \notag
\begin{pmatrix}
\ell & \ell' & L \\
-s   &  s  & 0
\end{pmatrix}. 
\end{align}
Taking the power spectrum of Equation (\ref{eq:b_lens}), we obtain:
\begin{equation} \label{eq:clbb_len}
C^{\text{B},\text{len}}_{\ell} \, \approx \, C^{\text{B},\text{unl}}_{\ell} \, + \, \frac{1}{2\ell + 1} \sum_{\ell' L} \left| f_{\ell \ell' L} \right|^2 C^{\text{E},\text{unl}}_{\ell'} \, C^{\phi}_{L}.
\end{equation}

It is useful to gain an intuitive
understanding of which scales in E-mode polarization and the lensing potential
source the lensing B modes.
\begin{figure}[tp]
\centering
\includegraphics[width=\columnwidth]{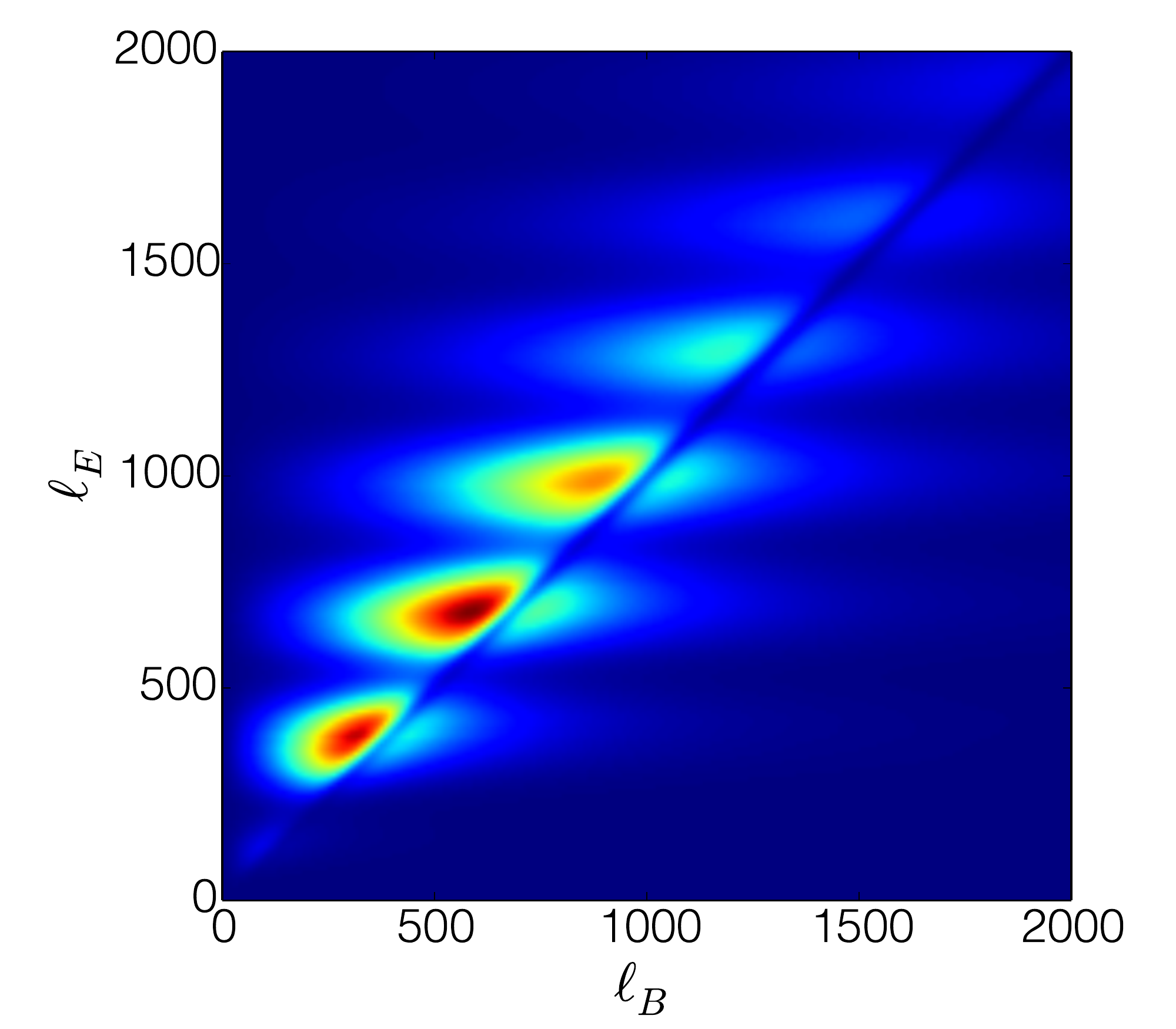}
\includegraphics[width=\columnwidth]{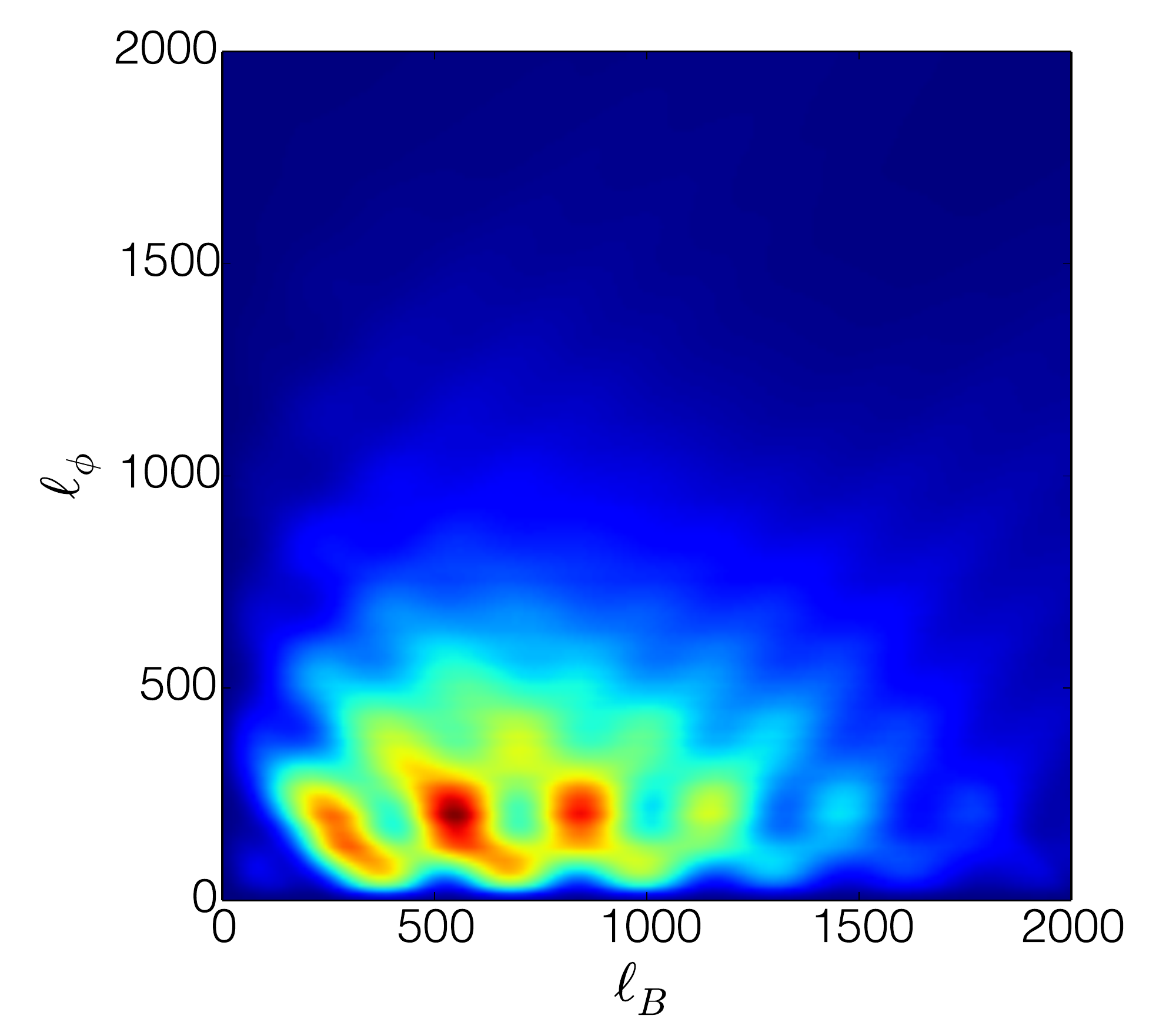}
\caption{Source terms for the lensed B modes.
We have plotted $\ell_B (\partial C_{\ell_B}^{B} / \partial C_{\ell_X}^{X})C_{\ell_X}^{X}$ where
$X \in \{ E, \phi \}$ such that the color along each column gives the fraction 
contribution per multipole to the lensed B modes from the corresponding source.
The lensing B modes at $l_B=600$, for example, are sourced largely by contributions 
from the lensing potential at $l_{\phi}\approx175$ and $l_E \approx 650$.
Delensing can be performed if one has measurements of these two fields
in the corresponding multipole ranges.}
\label{fig:which_scales}
\end{figure}
For this purpose, in Fig.~\ref{fig:which_scales}, we have plotted the kernel
for the lensing B-mode power spectrum, broken into contributions from 
$C_{\ell}^{\text{E},\text{unl}}$ and $C_L^{\phi}$.
It can be seen that in general the
important E modes contributing to a given $\ell$ in lensing B modes are
those from a scale that is just slightly smaller (larger in $\ell$).
Thus, to delens B modes up to $\ell \sim 500$ it is only required to
accurately measure E modes up to $\ell \sim 1000$, given a good map
of the gravitational potential.  
Obtaining accurate E-mode measurements on these scales is relatively 
straightforward: current ground-based CMB experiments such as 
SPTpol \citep{2012SPIE.8452E..1EA}, ACTpol \citep{2010SPIE.7741E..1SN}, and PolarBear \citep{2012SPIE.8452E..1CK} make sample-variance limited measurements
of the E-mode polarization on these scales. 

For the lensing potential, most of the lensing
B modes are generated by gravitational potential fluctuations on half-degree
scales or larger ($\ell \la 500$).
There are again a number of possible tracers for these modes.
CMB lensing measurements can
be used, although polarization-based lensing measurements
require high sensitivity and sensitivity to smaller angular scales than 
are required for large-scale tensor B-mode experiments (such as BICEP2).
Alternatively, astrophysical tracers could be used to estimate the
potential, provided they are sufficiently correlated with the lensing potential. 

The idea behind delensing is to build a noisy estimate of the lensing B modes using Wiener-filtered E modes and lensing potential $\phi$. This estimate is then subtracted from the perfectly reconstructed B modes, leaving as a difference the residual B modes:
\begin{align} \label{eq:clbb_res}
C^{\text{B},\text{res}}_{\ell} \, = & \;  \frac{1}{2\ell + 1} \sum_{\ell_1 \ell_2} \left| f_{\ell \, \ell_1 \ell_2} \right|^2  \Bigg[ C^{\text{E},\text{len}}_{\ell_1} \, C^{\phi}_{\ell_2}   \\
&- \Bigg( \frac{ \big( C^{\text{E},\text{len}}_{\ell_1} \big)^2 }{C^{\text{E},\text{len}}_{\ell_1} + N^{\text{E}}_{\ell_1}}  \Bigg) \Bigg( \frac{ \big( C^{\phi}_{\ell_2} \big)^2 }{C^{\phi}_{\ell_2} + N^{\phi}_{\ell_2}} \Bigg)  \Bigg]. \notag 
\end{align}
A formal derivation of this expression can be found in \citet{Smith:2008an}.
In writing Equation (\ref{eq:clbb_res}), we have substituted the \emph{unlensed} E-mode power spectrum coming from the derivation of Equation (\ref{eq:clbb_len}) by the \emph{lensed} E-mode power spectrum, which has been shown by \citet{Lewis:2011dfg} to produce an estimated lensed B-mode power spectrum effectively accurate to higher order.
The experimental noise power spectrum can be written as \citep{Knox:1995dq}:
\begin{equation}\label{eq:noise_pspec}
N^{E}_{\ell} \, = \, N^{B}_{\ell} \, = \, \Delta^2_P \, \exp \left ( \frac{\ell \left(\ell+1\right)\theta^2_{\text{FWHM}}}{8 \ln 2}  \right ),
\end{equation}
where $\Delta_P$ is the polarization pixel noise in [$\mu$K radian] and $\theta_{\text{FWHM}}$ is the FWHM of the beam, assuming this beam is Gaussian. We do not consider an additional foreground noise power spectrum.

The lensing reconstruction noise power spectrum $N^{\phi}_{\ell}$ can be obtained through the prevalent quadratic estimator method \citep{Hu:2000ee}. This technique builds an estimate of the lensing potential from the optimally filtered off-diagonal correlations between two lensed CMB fields $X, X'  \in \{ T, E, B \}$.
We use the EB estimator which yields the best signal to noise ratio at high experimental sensitivities. The resulting quadratic lens reconstruction noise level can be expressed as \citep{Okamoto:2003zw}:
\begin{align} \label{eq:nlpp_eb}
N^{\phi}_{\ell} \, = \;  &\Bigg[  \frac{1}{2\ell + 1} \sum_{\ell_1 \ell_2} \left| f^{\text{EB}}_{\ell \, \ell_1 \ell_2} \right|^2   \\
&\times  \Bigg( \frac{1}{C^{\text{B},\text{len}}_{\ell_1} + N^{\text{B}}_{\ell_1}}  \Bigg)   \Bigg( \frac{ \big( C^{\text{E},\text{len}}_{\ell_2} \big)^2 }{C^{\text{E},\text{len}}_{\ell_2} + N^{\text{E}}_{\ell_2}}  \Bigg) \Bigg]^{-1}. \notag
\end{align}
We used the fast real-space algorithm proposed by \cite{Smith:2012ghy} to compute the geometrical factors described in Equations (\ref{eq:f_coupling}) and (\ref{eq:F_coupling}) necessary for the numerical implementation of Equations (\ref{eq:clbb_res}) and (\ref{eq:nlpp_eb}).

It can be seen from Equation (\ref{eq:nlpp_eb}) that the lensing B modes act as a contaminant for the lensing reconstruction process.
Substituting the signal+noise lensed B modes of Equation (\ref{eq:nlpp_eb}) by delensed B modes, defined in this forecast as
\begin{equation} \label{eq:clbb_tot}
C^{\text{B},\text{del}}_{\ell} \, \approx \, C^{\text{B},\text{unl}}_{\ell} + C^{\text{B},\text{res}}_{\ell} + N^{\text{B}}_{\ell},
\end{equation}
one can estimate the lensing potential with higher signal-to-noise and then produce lower residual B modes.
Repeating these steps until convergence of $C^{\text{B},\text{del}}_{\ell}$ is referred to as iterative delensing \citep{Seljak:2003pn, Smith:2008an}. Quadratic delensing will indicate the use of a delensed power spectrum $C^{\text{B},\text{del}}_{\ell}$ computed through only one iteration.

Lensing reconstruction can also be achieved using large-scale structure \citep{Smith:2012ghy}.
We explore here the possibility of estimating the lensing potential by using its cross-correlation with the CIB, its best known tracer (\citealt{Hanson:2013hsb}; \citealt{Holder:2013hqu}; \citealt{Ade:2013aro}; \citealt{Ade:2013hjl}). 
In this work, 
we assume that high signal-to-noise CIB measurements are available,
and that the cross-correlation between the CIB and the lensing potential can be well-approximated as flat with a constant correlation coefficient
\begin{equation}\label{eq:f_corr}
f_{\text{corr}} = \frac{C^{\text{CIB} \times \phi}_{\ell}}{\sqrt{C^{\phi}_{\ell} C^{\text{CIB}}_{\ell}}}.
\end{equation}
This simple approximation is quite accurate (see for example Fig.~13 of \citealt{Ade:2013aro}).
Including the Poisson noise component to the CIB power spectrum in Equation (\ref{eq:f_corr}) causes $f_{\text{corr}}$ to fall off on small scales, but
 on scales larger than $\ell \sim 500$ of interest for delensing, the flatness of the correlation factor is not substantially affected. 
Although recent measurements have shown a correlation at the $f_{\text{corr}}=0.8$ level, we will pessimistically also consider delensing with correlation coefficients as low as $f_{\text{corr}}=0.4$.
This could effectively be the case, for example, in a region with comparable foreground and CIB power.
The introduction of this correlation coefficient results in a simple expression for the lensing estimate signal+noise power spectrum used in Equation \eqref{eq:clbb_res};
\begin{equation}
C^{\phi}_{\ell} + N^{\phi}_{\ell} = C^{\phi}_{\ell} / f^2_{\text{corr}}.
\end{equation}

Assuming Gaussianity of the likelihood function $\mathcal{L}(\mathbf{d} | \boldsymbol \theta)$ which gives the 
probability distribution of a cosmological model $\boldsymbol \theta$ given some set of independent observations $\mathbf{d}$, we use the Fisher matrix formalism to compute forecasted errors on the tensor tilt. For the present analysis, the data covariance matrix will be reduced to the B-mode power spectrum since the constraining power on cosmological parameters such as $r$ and $n_T$ comes mainly from the B-mode signal.
Each element of the Fisher matrix reduces to \citep{Jungman:1995av}
\begin{equation} \label{eq:fisher_ij}
F_{ij} =   \sum^{\ell_{\text{max}}}_{\ell_{\text{min}}} \frac{\big( C^{\text{B}}_{\ell} \big)_{\!, \, i}  \big( C^{\text{B}}_{\ell} \big)_{\!, \, j}}{ \left( \delta C^{\text{B}}_{\ell} \right)^2},  
\end{equation}
where the expected 1$\sigma$ error on the measurement of the B-modes power spectrum is 
\begin{equation} \label{eq:del_clbb}
\delta C^{\text{B}}_{\ell} = \sqrt{\frac{2}{(2\ell +1) f_{\text{sky}}}} C^{\text{B}}_{\ell}.
\end{equation}
The marginalized error on a given parameter of the model corresponds to 
$\sigma(\theta_i) = \sqrt{\left(F^{-1} \right)_{ii}  } $.

\section{Constraints on Tensor Tilt}
\label{sec:results}

\begin{figure}[tp]
\centering
\includegraphics[scale=0.5]{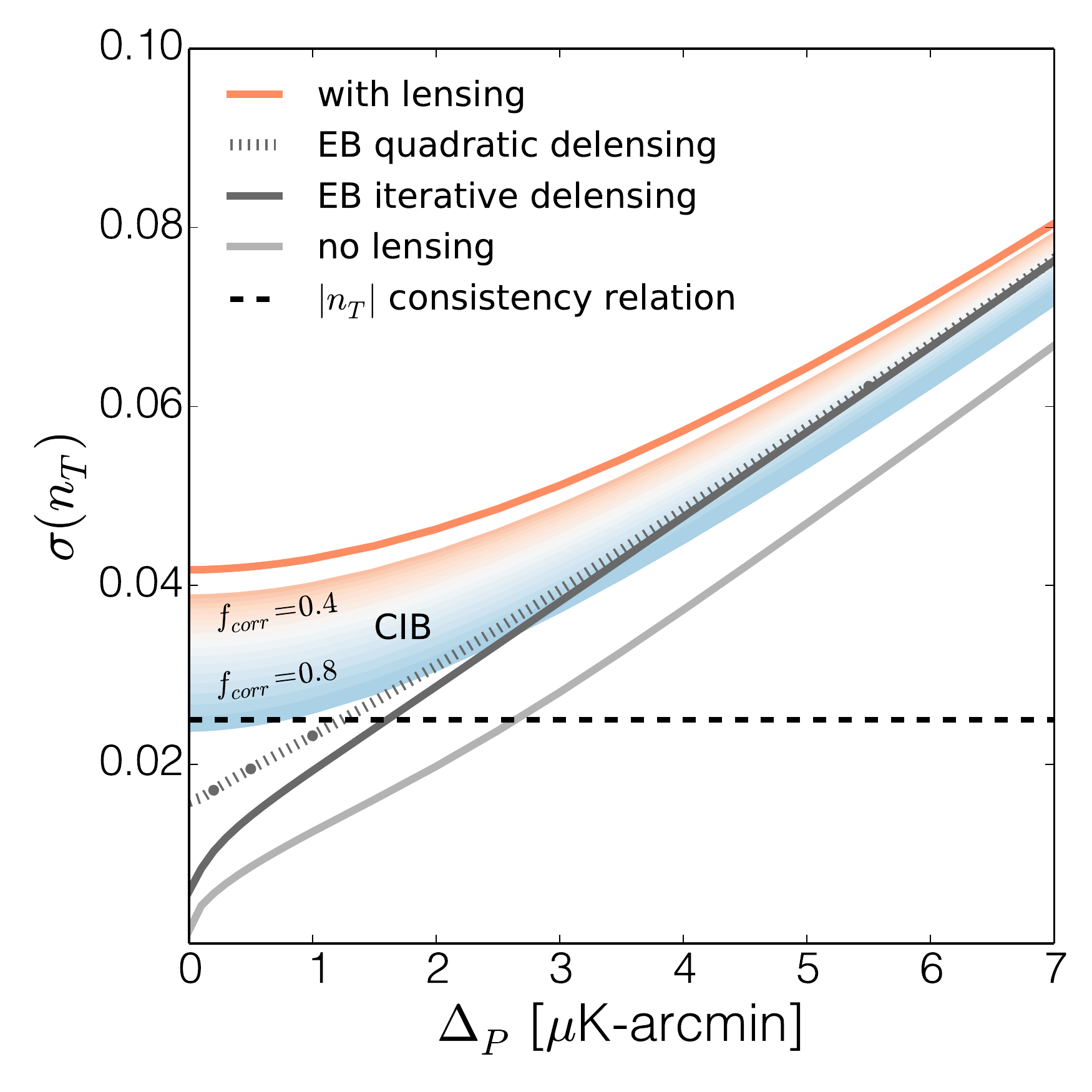}
\includegraphics[scale=0.5]{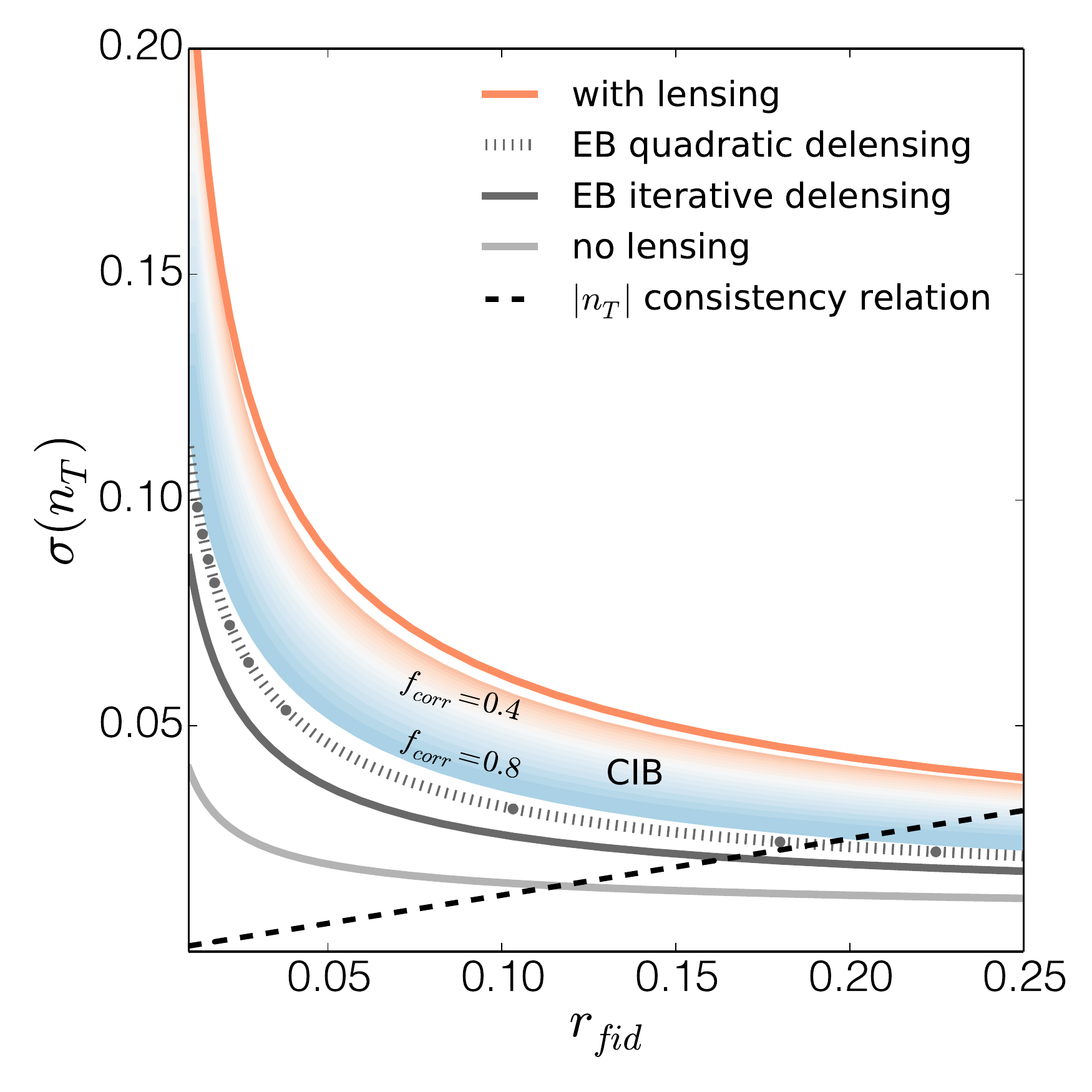}
\caption{
\emph{Top panel:}
forecasted constraint on the tensor tilt $n_T$ marginalized over $r$ for $r_{fid}=0.2$ and varying polarization sensitivity.
\emph{Bottom panel:}
forecasted constraint on the tensor tilt $n_T$ marginalized over $r$ for $\Delta_P$=1 $\mu$K arcmin and varying fiducial values of $r$. 
In both panels the black dashed line shows the absolute value of the tensor tilt given it is related to the fiducial value of $r$ through the consistency relation $n_T=-r/8$. The sky coverage has been fixed to $f_{\text{sky}}=0.5$ and the beamsize to 4 arcmin in both cases.
}
\label{fig:sigma_nt}
\end{figure}

We have computed the projected constraint on the tensor tilt assuming different delensing scenarios; results are shown in Fig.~\ref{fig:sigma_nt}.
The two limiting cases correspond to a situation where no lensing B modes are removed (solid orange curves) and where the total B modes include no lensing B modes (solid light gray curves). The solid and dotted dark gray curves correspond to quadratic and iterative CMB delensing and the shaded regions correspond to delensing using the CIB, where the correlation factor $f_{\text{corr}}$ is comprised between 0.4 and 0.8. 
For simplicity we will consider that the E-mode estimate and the lensing potential estimate are obtained through the same CMB experiment when considering CMB delensing, which makes the noise power spectra in Equations (\ref{eq:clbb_res}) and (\ref{eq:nlpp_eb}) parametrized by the same beam width and sensitivity. This forecast could be expanded to the case where a large-scale mission is used for the measurement of the E modes and a higher resolution experiment is used for estimating the lensing potential.

The beam width $\theta_{\text{FWHM}}$ has been fixed to 4 arcmin in both panels; it has been shown that beam size is not an important factor \citep{Boyle:2014kba, Dodelson:2014exa, Wu:2014hta}.
We have included modes between $l_{\text{min}} = 10$ and $l_{\text{max}} = 3000$ in the Fisher matrix calculation, although the large scale cutoff value does not affect significantly the projected constraints on $n_T$  \citep{Boyle:2014kba, Dodelson:2014exa}.
The sky coverage has been held to $f_{\text{sky}}=0.5$, knowing that for a high resolution experiment the results scale with $f_{\text{sky}}$ \citep{Dodelson:2014exa}.

The top panel of Fig.~\ref{fig:sigma_nt} shows the forecasted error on $n_T$ as a function of the polarization experimental noise level. As $\Delta_P$ approaches $0$, no fundamental floor due to the iterative delensing procedure is found, in conformity with previous works \citep{Seljak:2003pn, Smith:2012ghy, Wu:2014hta}.
Better than linear improvement on $\sigma(n_T)$ is observed as polarization noise drops below $\Delta_P$=1 $\mu$K arcmin \citep{Caligiuri:2014sla}.
This noise level also represents the sensitivity required to probe the tensor tilt at a level that is interesting for testing inflationary models; around $\Delta_P$=1 $\mu$K arcmin, $\sigma(n_T)$ falls below the dashed line corresponding to the consistency relation.
Given the most optimistic case of correlation level between the CMB and the CIB, it appears that
CIB-based delensing will not be sufficient for measuring the tensor tilt. However, for the current next generation of CMB polarization experiments, CIB-based delensing will be of comparable utility.

The error on $n_T$ depends on the value of $r$ with which the B-modes power spectrum $C_{\ell}^B$ in Equations (\ref{eq:fisher_ij}) and (\ref{eq:del_clbb}) is computed. This dependence is plotted in the bottom panel of Fig.~\ref{fig:sigma_nt}, for a fixed pixel noise of $\Delta_P$=1 $\mu$K arcmin.
The dashed black line shows the absolute value of $n_T$, given that $n_T$ and $r$ are related by the consistency relation. It can be seen that going from $r=0.2$ down to $r=0.1$ hinders the ability to distinguish between consistency relation and other models \citep{Caligiuri:2014sla, Dodelson:2014exa}. 

\section{Conclusions}
\label{sec:conclusions}

We have tested the ability of different delensing techniques, namely EB quadratic delensing, EB iterative delensing and CIB-based delensing to constrain the tensor tilt.
For low CMB noise levels, CMB-based delensing would be able to probe inflation with a constraint on $n_T$ given the initial BICEP2 signal is entirely of cosmological origin.
This would require sensitivities on the order of 1 $\mu$K arcmin in polarized noise over roughly half the sky.
While ambitious, this is exactly the scale that is being considered
as Stage-IV CMB experiments like CMB-S4 (Abazajian et al. 2013).
CIB delensing cannot constrain $n_T$ to a level interesting for probing inflation, although it is competitive with near-future CMB delensing strategies.

\section*{Acknowledgments} 

While this work was being achieved, the preprint of \cite{Boyle:2014kba} was released, which has strong overlaps with the results presented in this paper. 
G. S. wishes to thank Ryan Keisler and Elisa G. M. Ferreira for useful discussions and comments. 
We acknowledge the use of Kendrick Smith's implementation of the Gauss-Legendre quadrature method and of the Wigner-d matrix.
G. H. acknowledges funding from the Canadian Institute for Advanced Research, the National Sciences and Engineering Research Council of Canada and Canada Research Chairs program.
D. H. is supported by the Lorne Trottier Chair in Astrophysics and Cosmology at McGill as well as a CITA National Fellowship.
G. S. acknowledges support from the Fonds de recherche du Qu\'ebec - Nature et technologies.

\end{document}